# Current-Induced Nanomagnet Dynamics for Magnetic Fields Perpendicular to the Sample Plane

S. I. Kiselev, J. C. Sankey, I. N. Krivorotov, N. C. Emley, M. Rinkoski, C. Perez, R. A. Buhrman and D. C. Ralph
*Cornell University, Ithaca NY, 14853*
(Dated March 3, 2004)

We present electrical measurements of high-frequency magnetic dynamics excited by spin-polarized currents in Co/Cu/Ni$_{80}$Fe$_{20}$ nanopillar devices, with a magnetic field applied perpendicular to the sample layers. As a function of current and magnetic field, the dynamical phase diagram contains several distinguishable precessional modes and also static magnetic states. Using detailed comparisons with numerical simulations, we provide rigorous tests of the theory of spin-transfer torques.

PACS numbers: 72.25. Ba, 73.40.-c, 75.40. Gb

Spin-polarized electron currents can apply torques to magnets by the direct transfer of spin angular momentum.[1,2] Experiments have shown that DC currents $I$ can drive reversible switching of a nanomagnet between static orientations,[3-6] or can excite the magnetization into steady-state dynamical modes.[7-10,4,5] Previously, the types of dynamical states caused by spin transfer for isolated nanomagnets have been determined only for magnetic fields $H$ applied in the plane of the thin-film magnets.[11] A perpendicular-field study has been performed for excitations of domains strongly coupled within an extended magnetic film, but this coupling inhibited direct comparisons with theory.[12] Here we describe detailed measurements of the dynamics of isolated nanomagnets in perpendicular fields. With this geometry we are able to tune the orientation of the magnetic moment continuously out of the sample plane, enabling us to characterize several dynamical modes not observed previously. The expanded phase diagram provides a stringent test for theories of spin-transfer torques.[13-17]

We study devices fabricated by first sputtering a multilayer consisting of 80 nm Cu / 40 nm Co / 10 nm Cu / 3 nm Ni$_{80}$Fe$_{20}$ / 2 nm Cu / 30 nm Pt. The layers are ion-milled to form a pillar structure with a 130 nm × 70 nm elliptical cross section, planarized with silicon oxide, and contacted on top by a Cu electrode [11] (inset, Fig. 1(a)). The Co layer is only partially patterned so that it can be reoriented easily by $H$, but its thickness and its coupling to an extended film make it relatively insensitive to current-induced torques. We employ a permalloy (Py = Ni$_{80}$Fe$_{20}$) "free layer" following [10] to minimize crystalline anisotropies. Positive $I$ is defined so that electrons travel from Py to Co.

The volume of the Py free layer is small enough that it is superparamagnetic at room temperature. For $H$ applied in the plane of the sample, the room-temperature dynamical phase diagram (Fig. 1(a)) for the free layer is similar to that measured

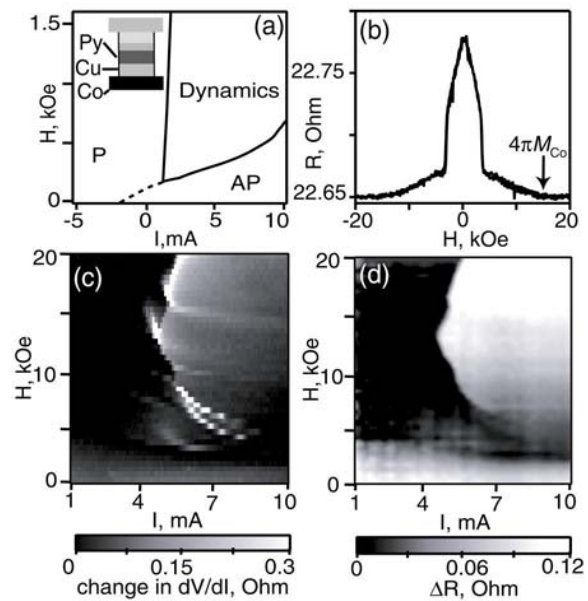

FIG. 1 (a) Measured room-temperature dynamical stability diagram for in-plane $H$. The dashed line separates static states with the free-layer moment parallel (P) and antiparallel (AP) to the fixed layer. Solid lines outline the region of microwave-frequency dynamics. Inset: Schematic of the sample geometry. (b) Magnetoresistance near $I=0$ for $H$ perpendicular to the sample layers. The measurement includes a lead resistance of ~ 20 Ω. (c) Variations in $dV/dI$ and (d) in $R=V/I$ for perpendicular $H$ as a function of $I$.

previously for Co samples except that (*i*) at low $H$ because of superparamagnetism there is a simple crossover from parallel alignment (P) relative to the fixed layer to antiparallel alignment (AP) rather than a region of P/AP bistability, and (*ii*) for the Py samples we do not observe the region "W" described in ref. [11], which might be associated with the breakdown of single-domain dynamics. The two magnetic layers have a dipole coupling equivalent to a field $H_d$ = 180 ± 10 Oe on the free layer. At 4.2 K, the free layer is no longer superparamagnetic; the low-temperature currents and fields for switching the Py layer are $I_c^{P \to AP} - I_c^{AP \to P}$ = 2.5 mA and $H_c$ = 300 ± 20 Oe (so that the within-plane anisotropy is $H_{an} \approx H_c - H_d =$ 120 ± 20 Oe). The effective perpendicular demagnetization fields are $4\pi M_{Co} \sim 15$ kOe and $4\pi M_{Py} \sim 5$ kOe (determined below).

The orientation of $H$ for the rest of the measurements we report will be approximately perpendicular to the sample plane, with 1-2° of tilt toward the easy axis of the free layer so as to control the in-plane orientation of the Co moment. We have characterized the samples using conventional low-frequency resistance measurements at room temperature (Figs. 1(b)-(d)), with results similar to those reported for Co/Cu/Co samples at 4.2 K.[18] From Fig 1(b), we can identify $4\pi M_{Co} \sim 15$ kOe. As $I$ is ramped to positive values, we observe peaks in the differential resistance $dV/dI$ (Fig. 1(c)), associated with sudden increases in the DC resistance $R = V/I$ of the sample (Fig. 1(d)). As noted previously,[18] for sufficiently large $H$ ($H > 4\pi M_{Co}$) there is a single transition with a resistance change equal to the full room-temperature difference between P and AP magnetic orientations $\Delta R = 0.12$ Ω. For smaller $H$, $\Delta R < 0.12$ Ω, and sometimes there is more than one transition as a function of $I$. We see no hysteresis as a function of the direction of current sweep for any value of $H$ at room temperature.

To determine the types of magnetic dynamics associated with these resistance changes, we employ the technique described in [11]. A DC current is injected to drive the magnetic dynamics, causing the resistance of the sample to change at microwave frequencies due to the changing relative angle between the moments of the magnetic layers. This results in a voltage oscillation, whose spectrum is recorded using a heterodyne spectrometer circuit.[11] We examine the frequency range 0.5-40 GHz by making separate measurements using two sets of amplifiers with bandwidths 0.5-18 GHz and 18-40 GHz. Each spectrum displayed in this Letter is obtained by subtracting the Johnson noise background from the raw data and then dividing the result by $I^2$, so that the final result reflects the magnitude of the resistance oscillations.

As a function of increasing $I$ at fixed $H$, the measured microwave signals initially appear with small amplitude, and subsequently exhibit a variety of changes that we can associate with transitions between dynamical states. In order to determine the sample parameters to be used in simulations, we first consider the frequency of signals measured at small $I$, when the dynamics first become visible (Fig. 2(a)). The linear dependence of $f$ for large $H$ agrees with the Kittel formula for small-angle precession of a moment about an axis perpendicular to the sample plane [19]:

$$f = g\mu_B (H - 4\pi M_{Py})/2\pi, \qquad (1)$$

where $\mu_B$ is the Bohr magneton. A linear fit for $H > 6.5$ kOe (red line) gives $g = 2.23 \pm 0.02$ and $4\pi M_{Py} = 5.1 \pm 0.1$

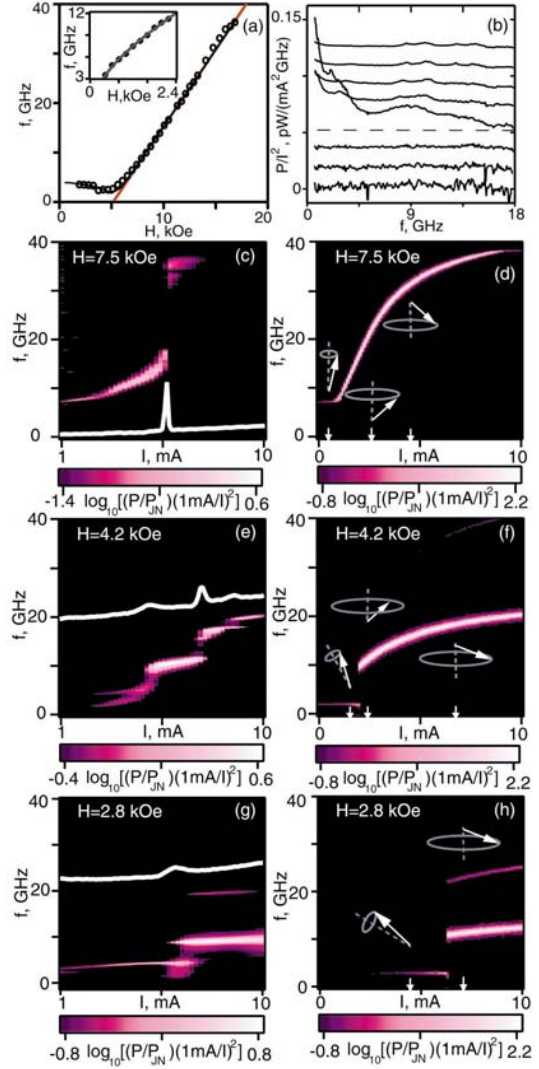

FIG. 2 (color) (a) Magnetic field dependence of the measured small-amplitude signal frequency for perpendicular $H$ (circles) compared to numerical simulations of the LLG equation (black curve). The red line is a linear fit to Eq. (1). Inset: $H$ dependence of the small-amplitude signal frequency for in-plane $H$, with a fit to the Kittel formula. (b) Microwave spectra for $H = 18$ kOe perpendicular to the sample plane, for $I = 2.2$-$9.2$ mA, in increments of 1 mA (offset vertically). (c,e,g) Color scale: Measured microwave spectra as a function of $I$ at selected values of $H$ perpendicular to the sample plane. The spectra are normalized by the magnitude of room-temperature Johnson noise, $P_{JN}$, and have been divided by $I^2$, as discussed in the text. White lines show $dV/dI$, plotted using different vertical scales. (d,f,h) Spectra predicted by the simulation described in the text. The diagrams illustrate the dynamical modes of the free layer within the simulations, for the values of $I$ marked by arrows. The vertical direction is normal to the sample plane.

kOe. Similar parameters also fit the small-angle precession frequencies for $H$ applied in plane (Fig. 2(a) inset); by fixing $g = 2.23$, a fit to the Kittel formula for this case [19] gives $4\pi M_{Py} = 4.4 \pm 0.2$ kOe. Over the full range of $H$ shown in Fig. 2(a), the measured frequencies agree well with the values expected for small-angle precession about the equilibrium angle of the Py moment, as determined from Landau-Lifshitz-Gilbert (LLG) simulations which treat the Py layer as a single magnetic domain.[1] The black curve in Fig. 2(a) gives $f$ predicted by this simulation with no fitting parameters, using the values $4\pi M_{Py}$ = 5 kOe, $g$ =2.23, $H_d$=180 Oe, and $H_{an}$=120 Oe determined as described above.

The full dependence of the microwave spectra on $I$ is shown for selected values of $H$ in Fig. 2(b,c,e, and g). Different dynamical regimes are observed, depending on the magnitude of $H$ relative to the demagnetization fields $4\pi M_{Py}$ and $4\pi M_{Co}$. We will analyze these data in comparison with numerical LLG simulations of single-domain magnetic dynamics that include a term for the spin-transfer torque derived by Slonczewski[1] (Fig. 2(d, f, and h)). In the simulations, we use the fixed values for the parameters noted in the previous paragraph, along with $4\pi M_{Co}$ = 15 kOe. The only parameters not determined by independent measurements are the spin polarization of the current and the damping coefficient α, for which we use polarization = 0.3 and α = 0.012.[20] The types of dynamical modes we find are not sensitive to variations in the polarization or α. Room-temperature thermal effects are modeled by a fluctuating magnetic field following ref. [21].

In Fig. 2(b), we consider microwave spectra measured at $H$ = 18 kOe. This field is greater than $4\pi M$ for both magnetic layers, so that at $I$=0 both moments are aligned along $H$. We do not observe dynamics with any well-defined frequency for $H \geq 17$ kOe. The only microwave signal observed in this range of $H$ occurs in a very narrow region of current, corresponding to the peak in $dV/dI$, and consists of a broad spectrum with a low-frequency tail. Because any dynamical states that exist in the large-$H$ regime could possibly take the form of purely circular precession about the direction of the fixed layer moment, producing no resistance oscillations, we also searched for dynamical states by tilting $H$ approximately 10° from the sample normal to break circular symmetry. Away from the region of the peak in $dV/dI$, we observed only static states within our bandwidth for $H$ > 17.2 kOe. This supports the conclusion made in ref. [18] that for large perpendicular $H$ the free layer simply undergoes a transition from P to AP alignment over a narrow range of current. The absence of steady-state dynamical modes in this range of $H$ at room temperature agrees with the simulations, although the simulations suggest some hysteresis in the transition current, depending on the direction of current sweep, that we do not observe.[18]

Fig. 2(c) corresponds to $H$ = 7.5 kOe, in between $4\pi M_{Py}$ and $4\pi M_{Co}$. In the absence of $I$, the Py layer moment is therefore aligned with the field while the Co moment is tilted approximately 30° out of plane due to its larger demagnetization field. At $H$ = 7.5 kOe, a peak in the microwave signal is observed at $f$ = 7.5 GHz for $I$ as small as 0.8 mA, corresponding to small-angle precession of the free layer. Near $I$ = 2.6 mA, the dependence of $f$ on $I$ exhibits a kink, at which point the mean-square resistance oscillations grow by 50% and the width (FWHM) of the spectral peak broadens from 250 MHz (below 2.0 mA) to > 700 MHz. A similar transition is observed in the simulation, due to a change from small-angle precession to larger-angle dynamics (at $I$ = 1.0 mA in Fig. 2(d)). At larger currents, near $I$ = 5 mA in the experimental data (at the peak in $dV/dI$), there is a single large step in $f$ to a cluster of peaks near $f$ = 35 GHz. This step can be explained by a reversal of the free-layer's precession axis from nearly parallel to nearly antiparallel to $H$, so that the demagnetization field after the transition adds to $H$ rather than subtracting from it. This transition is abrupt in all samples measured, but in the simulations the reorientation of the free layer is predicted to occur continuously. Other features of the experiment that are not observed in the simulation are the existence of multiple peaks in the frequency spectrum following the transition to the antiparallel-to-$H$ precessional state and the disappearance of the precessional signal beyond 7.5 mA in Fig. 2(c).

Finally we consider $H$ = 4.2 kOe and 2.8 kOe (Fig. 2(e-h)), less than $4\pi M$ for both layers. For these fields, the kink in $f$ as a function of $I$ observed at 7.5 kOe is replaced by a step increase in $f$, at $I$ = 4.6 mA for $H$ = 4.2 kOe and $I$ = 5.8 mA for $H$ = 2.8 kOe. At this step a second-harmonic signal also becomes prominent. This behavior is reproduced nicely within the simulations, although at somewhat smaller values of $I$. The step can be identified with a reorientation of the free layer's precessional axis from its $I$=0 equilibrium angle to the applied field direction. At $H$ = 2.8 kOe the microwave power just prior to this transition is 3.2 pW/mA$^2$, in reasonable agreement with the simulated value 2.2 pW/mA$^2$, corresponding to a precessional angle of ~ 10°. However, there is a larger difference between the measured and simulated microwave powers following the step in $f$. Simulation predicts an integrated signal power of 77 pW/mA$^2$, while the measured signals are a factor of 6 smaller. The frequencies predicted by the simulations at large $I$ in Fig. 2(f,h) agree well with the measurements, although the simulations predict a continuous evolution of the precession frequency after the first step in $f$, while multiple steps in $f$ are seen in Fig. 2(e) beyond 5 mA, each associated with peaks in $dV/dI$.

The experimental results are summarized in Fig. 3(a), where we plot the positions of all of the steps in $f$, kinks in $f$, and peaks in $dV/dI$ that suggest transitions between dynamical modes. Filled symbols illustrate transitions that are observed in all samples. The frequency steps shown with open symbols are only qualitatively similar between samples (*e.g.*, those beyond 5 mA in Fig. 2(e)), with differences in the number of steps and the pattern of critical currents. The very large step in frequency seen at larger fields ($H$ = 7.5 kOe, Fig. 2(c)) is reproducible between samples. The abbreviations SLR and SHR denote static low-resistance and static high-resistance states.

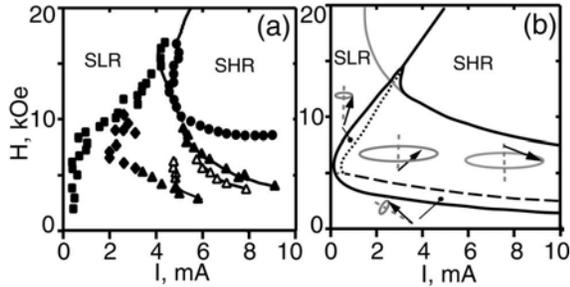

Fig. 3(a) Experimental room-temperature stability diagram for perpendicular *H*. Filled symbols indicate features that are reproducible between samples. Open symbols show transitions that can differ in number and position between samples. SLR (SHR) denotes a region of static low (high) resistance. Squares indicate the onset of small angle precession. Triangles denote the position of frequency steps. Diamonds show kinks in the dependence of *f* on *I*. Circles denote points where the microwave signal drops below measurable values. Lines indicate peaks in *dV/dI*. (b) Theoretical room-temperature stability diagram obtained by numerical solution of the LLG equation with the Slonczewski spin-torque term [1]. Solid lines separate SLR, SHR and dynamical states (precession angles > 2°). Dotted and dashed lines show the transition from small-angle precession about the *I*=0 equilibrium angle to large-angle precession about *H*, associated with a step in *f* (dashed line) or a kink in *f* vs. *I* (dotted line). Within the dynamical region, the precession angle is predicted to increase continuously as a function of *I*. At the 1 μs time scale in the simulations, transitions between the SLR and SHR states are hysteretic [18], with the boundary given by the black line for increasing *I* and the gray line for decreasing *I*.

The overall phase diagram predicted by the single-domain simulations for *H* perpendicular to the sample plane is summarized in Fig. 3(b). Qualitatively, there is good correspondence between the model and our data (Fig. 3(a)) for the relative positions of the static high and low-resistance states and the dynamical region. The simulation also gives a good account of the transition between small-angle precession about the equilibrium (*I*=0) free-layer orientation and the regime of larger-angle dynamics about the direction of *H* (the dotted and dashed lines in Fig. 3(b)). However, the model does not predict the other steps in *f* observed within the dynamical regime.

The differences that exist between the measurements and the simulations could be due to several factors. The peaks in *dV/dI*, associated small steps in *f*, that are not reproducible between samples are reminiscent of signals analyzed in ref. [22] and ascribed there to coupling with lattice vibrations. Similar steps in *f* were observed in [12]. Because these transitions do differ between samples, we think it is more likely that they are due to irregularities in the magnetic anisotropy function describing each sample, which might act to stabilize some trajectories for the free-layer moment, with jumps between trajectories. However, because the large step in *f* to the antiparallel-to-*H* precessional state (Fig. 2(c)) is reproducible between samples, a more fundamental mechanism may be required to explain the discontinuous nature of this transition. It is possible that an angular dependence stronger than we have assumed for either the strength of the spin-transfer torque or for the damping term in the LLG equation could destabilize the continuous evolution of the dynamics predicted in Fig. 2(d) and produce instead a sudden transition to precession antiparallel to *H*. Stronger angular dependences for both the spin-transfer torque [23] and the damping [24] have been proposed, and the consequences on the dynamics should be studied in detail. It is also likely that some large-amplitude dynamical modes may require descriptions beyond our single-domain approximation.

In summary, we measured microwave-frequency dynamics in Co/Cu/Py nanopillar devices excited by a DC spin-polarized current, with a magnetic field perpendicular to the sample plane. Several different dynamical modes can be excited, with properties that are generally in accord with predictions of single-domain simulations. This agreement provides strong support for the framework of the Slonczewski description of spin-transfer torques.[1] However, the existence of some differences between measurements and simulations, particularly the observation of discontinuous transitions between dynamical states that are predicted to be continuous, suggests that descriptions of spin-transfer-driven dynamics at large excitation angles are not yet complete.

We thank A. D. Kent for discussions. We acknowledge support from the NSF/NSEC program through the Cornell Center for Nanoscale Systems, from DARPA through Motorola, and from the Army Research Office. We also acknowledge use of the NSF-supported Cornell Nanofabrication Facility/NNIN.